\input{aipcheck}

\documentclass[
    ,final            
  ]
  {aipproc}

\layoutstyle{6x9}

\begin{document}

\title{The nature of the present}

\classification{01.70.+w, 04.20.-q, 04.20.Gz}
              
\keywords      {Philosophy of science, general relativity, space-time}

\author{Gustavo E. Romero}{address={Instituto Argentino de Radioastronomía (CCT- La Plata, CONICET), C.C.5, 1894 Villa Elisa, Buenos Aires, Argentina.},
  altaddress={Facultad de Ciencias Astronómicas y Geofísicas, Universidad Nacional de La Plata, paseo del Bosque, 1900, La Plata, Argentina.}}


\begin{abstract}
 The feeling of a moving present or `now' seems to form part of our most basic perceptions about reality. Such a present, however, is not reflected in any of our theories of the physical world. In this short note I argue for a tenseless view of time, where what we call `the present' is just an emergent secondary quality arising from the interaction of perceiving self-conscious individuals with their environment. I maintain that there is no flow of time, but just an ordered system of events.

\end{abstract}

\maketitle


\section{Introduction}

Time has always puzzle philosophers and scientists alike. Traditionally, there are two broad views about the nature of time. These views are usually called the ``tensed'' and the ``tenseless'' views, or, for simplicity, the A and B views of time. For an A-oriented person, only present things exist. There are many varieties of this ontological position: presentism, becoming theory, primitive tenses, branching universe theory, and so on. All of them distinguish the present in some way. In particular, presentism is the doctrine that it is always the case that, for every $x$, $x$ is present. The quantification in this definition is unrestricted, it ranges over all existents. In order to make this definition meaningful, the presentist must provide a specification of the term ``present''. A standard definition is:\\

{\bf Present}: The mereological sum of all objects with null temporal distance \cite{Crisp-03-Oxford}.\\

Since the mereological sum of objects is always an object, we can infer that for a presentist the present is an object, i.e. and individual with some properties. 

A B-oriented person will consider all this as pure nonsense. She will maintain that past, present and future `equally' exist. For her, the fundamental temporal properties are relations of `earlier than', `later than' and `simultaneous with'. These are relations between events. There is no distinguished present in any absolute sense. The present is {\sl not} an object. Then, it cannot move, since only objects can move with respect to each other. There is no objective `flow' or passage of time.  

What is, then, the present in this view? My aim, in this short note, is to answer this question from a B-perspective.

\section{Shoot the presentist}

The Englishman John McTaggart Ellis McTaggart (1866-1925) presented a disproof of presentism in his famous paper {\sl Unreality of Time} \cite{McT-1908}. He reasoned as follows.
\begin{enumerate}
\item There is no time without change.
\item If time passes, events should change respect to the properties of pastness, presentness, and futureness.
\item A given event, then, should be able to be in absolute sense, past, present and future. 
\item These properties exclude each other.
\end{enumerate}
Then: Events do not pass, just are. \\

There is no passage of time. There is no moving present. The mere idea of a flowing time simply does not make any sense. An additional problem is that if time flows, it should move with respect to something. If we say that there is a super-time with respect to time flows, then we shall need a super-super-time for this super-time, and we shall have an infinite regress. In addition, there is no flow without a rate of flow. At what rate does time go by? The answer 1 sec per sec is meaningless. It is like saying that a road extends along a distance  of one km per each km that it extends! 

On the physical side, the theory of special relativity seems not to be friendly to the idea of an absolute present, at least in its usual Minkowskian 4-dimensional interpretation (for arguments against presentism from general relativity see the paper by Romero and P\'erez in this volume). Special relativity is the theory of moving bodies formulated by Albert Einstein in 1905 \cite{Einstein}. It postulates the Lorentz-invariance of all physical law statements that hold in a special type of reference systems, called {\it inertial frames}. Hence the `restricted' or `special' character of the theory. The equations of Maxwell electrodynamics are Lorentz-invariant, but those of classical mechanics are not. When classical mechanics is revised to accommodate invariance under Lorentz transformations between inertial reference frames, several modifications appear. The most notorious is the impossibility of defining an absolute simultaneity relation between events. The simultaneity relation results to be frame-dependent. Then, some events can be future events in some reference system, and present or past in another system. Since what exists cannot depend on the reference frame adopted for the description of nature, it is concluded  that past, present, and future events exist. Then, presentism, the doctrine that only present exists, is false.

The presentist or A-theorist of time might find a way around this argument adopting a different (purely Lorentzian) interpretation of the theory \cite{Crisp-07}. The problems of this approach has been discussed at length by Saunders \cite{Saunders}, and I shall not insist on the topic here.

Said all that, yet, we all have a kind of feeling that ``our time is running out''. Where does this feeling come from?              
 
\section{When is `now'?}

If we assume that the present is an instant of time instead of a thing, then the question of ``which instant is present?'' follows. One possible answer is ``now''. But...when is `now'?

`Now', like `here', is an indexical word. To say that I exist now gives no information on when I exist. Similarly, to say that I am here, gives no information on where I am. There is no particular moment of time defined as an absolute now. 

I maintain that `nowness' and `hereness' emerge from the existence of perceiving self-conscious beings in a certain environment. What these beings perceive is {\sl not} time, but changes in things. Similarly, they do not perceive space, but spatial relations among things. In particular, we do not perceive the passage of time. We perceive how our brain changes. I submit that there is no present {\sl per se}, in the same way that there is no smell, no pain, no joy, no beauty, no noise, no secondary qualities at all without sentient beings. What we call ``the present'' is not in the world. It emerges from our interaction with the world. 

We group various experienced inputs together as present; we are tempted to think that this grouping is done by the world, not by us. But this is just delusional.  
I maintain that tenses aren't needed and in fact aren't wanted by the natural sciences. This idea is clearly expressed by Poeppel on the basis of neurological research \cite{Poeppel}:

\begin{quotation}

[...] our brain furnishes an integrative mechanism that shapes sequences of events to unitary forms...that which is integrated is the unique content of consciousness which seems to us present. The integration, which itself objectively extends over time, is thus the basis of our experiencing a thing as present.

[...] The now, the subjective present, is nothing independently; rather it is an attribute of the content of consciousness. Every object of consciousness is necessarily always now - hence the feeling of nowness.
\end{quotation}

The perception of motion gives an additional argument against the idea that the present is an instant of time. According to Le Poidevin \cite{Poidevin}: 

\begin{enumerate}

\item What we perceive, we perceive as present.
\item We perceive motion.
\item Motion occurs over an interval.

\end{enumerate}

Therefore: What we perceive as present occurs over an interval. \\

Any tentative definition of `present' compatible with modern neuro-biological science must take into account the role of the perceiving and sentient individual. In the next section I shall offer some provisional definitions that meet this requirement and distinguish among the different meanings in which the word `present' is used.

\section{Defining the present}

Physical events are ordered by the relation `earlier than' or `later than', and `simultaneous with' \cite{Gru-73-Rei}. There is no `now' or `present' in the representation of the physical laws. 
What we call `present' is not an intrinsic property of the events nor an instant of time, much less a moving thing. `Present' is a relation between a certain number of events and a self-conscious individual.\\

{\bf Present}: Class of all events simultaneous to a given brain event.\\

To every brain event there is a corresponding present. The individual, however, needs not to be aware of all events that form the present. The present, being a class of events, is an abstract object without any causal power.\\

{\bf Psychological present}: Class of local events that are causally connected to a given brain event. \\

Notice that from a biological point of view only local events are relevant. These events are those that directly trigger neuro-chemical reactions in the brain. Such events are located in the immediate causal past of the brain event. The psychological present is a conceptual construction of the brain, based on abstraction from events belonging to an equivalence class. The present, then, is not a thing or a change in a thing (an event).

William James introduced the concept of `specious present' as ``the short duration of which we are immediately and incessantly sensible'' \cite{James}. We re-elaborate this to get the following definition:\\ 

{\bf Specious present}: Length of the time-history of brain processes necessary to integrate all local events that are physically (causally) related to given brain event.\\

The specious present, then, being related to a brain process, can be different for different individuals equipped with different brains. The integration of the specious present can be performed in different ways, depending on the structure of the brain. It is even possible to imagine integration systems that can produce more than one specious present or even systems that might `recall' the future \cite{Hartle}. If biological evolution has not produced such systems seems to be a consequence of the existence of space-time asymmetric boundary conditions that introduce a preferred direction for the occurrence of processes \cite{Romero-Perez}.

Finally, we can introduce a {\sl physical present}.\\

{\bf Physical present}: Class of events that belong to a space-like hypersurface in a smooth and continuous foliation of a time-orientable space-time. \\

Since in the manifold model of space-time every event is represented by an element of the manifold, the introduction of this class does not signal a special time identified with `now'. Every space-like hypersurface corresponds to a different time and none of them is an absolute present ``moving'' into the future. Actually, naming `the future' to a set of surfaces in the direction opposite to the so-called Bing Bang is purely conventional.

\section{Final remarks}

We have distinguished three different types of present: psychological, physical, and specious. The former two are classes of events, hence they are concepts. The latter is not an instant of time but an interval in space-time associated with the world history of a sentient individual.

In any case, the present does not flow or move. Only material individuals (and their brains, if they have one) can change. Becoming is not a property of physical events, but of the consciousness  of the events. We call `becoming' to the series of states of consciousness associated with a certain string of physical events. Events do not become. Events just {\em are}.



\begin{theacknowledgments}
I thank Daniela P\'erez and Felipe Tovar Falciano for insightful comments. This work was supported by CONICET (PIP 0078).
\end{theacknowledgments}





\begin{thebibliography}{}

\bibitem[\protect\citeauthoryear{Crisp}{2003}]{Crisp-03-Oxford}
Crisp, T. (2003).
\newblock Presentism. (In M. J. Loux \& D. W. Zimmerman (Eds.), {\em The Oxford Handbook of Methaphysics} (pp. 211-245). Oxford: Oxford University Press).

\bibitem[\protect\citeauthoryear{mcTaggart}{1908}]{McT-1908}
McTaggart, J.M.E. (1908).
\newblock {Unreality of time}.
\newblock {\em Mind}, 17,  456-473.

\bibitem[\protect\citeauthoryear{Einstein}{1905}]{Einstein}
Einstein, A. (1905).
\newblock Zur Elektrodynamik bewegter K\"orper. {\em Annalen der Physik}, 17 (10), 891–921.

\bibitem[\protect\citeauthoryear{Crisp}{2007}]{Crisp-07}
Crisp, T. (2007).
\newblock Presentism, Eternalism and Relativity Physics. (In W. L. Craig \& Q. Smith (Eds.), {\em Einstein, Relativity and Absolute Simultaneity} (pp. 262-278). London: Routledge)

\bibitem[\protect\citeauthoryear{Sau}{2002}]{Saunders}
Saunders, S. (2002).
\newblock How relativity contradicts presentism. (In C. Callender (Ed.), {\em Time, Reality $\&$ Experience, Royal Institute of Philosophy, Supplement} (pp. 277-292). Cambridge, New York: Cambridge University Press)
\bibitem[\protect\citeauthoryear{Poeppel}{1978}]{Poeppel}
Poeppel, E. (1978).  
\newblock {Time perception}.
\newblock (In Richard Held et al. (eds.), {\em Handbook of Sensory Physiology, Vol. VIII: Perception} (pp. 713-729). Berlin: Springer-Verlag).

\bibitem[\protect\citeauthoryear{Le Poidevin}{2009}]{Poidevin}
Le Poidevin, R. (2009).  
\newblock {The experience and perception of time}.
\newblock {\em The Standfrod Encyclopedia of Philosophy}.
\newblock (http://plato.stanford.edu/entries/time-experience/).


\bibitem[\protect\citeauthoryear{Gru}{1973}]{Gru-73-Rei}
Gr\"unbaum, A. (1973).
\newblock {\em Philosophical Problems of Space and Time}.
\newblock (Dordrecht: Reidel).

\bibitem[\protect\citeauthoryear{Jam}{1893}]{James}
James, W. (1893). 
\newblock {\em The Principles of Psychology}. 
\newblock (New York: H. Holt and Company). 

\bibitem[\protect\citeauthoryear{Hartle}{2005}]{Hartle}
Hartle, J.B. (2005).
\newblock {The physics of now}.
\newblock {\em American Journal of Physics}, 73,  101-109.

\bibitem[\protect\citeauthoryear{Romero-Perez}{2011}]{Romero-Perez}
Romero, G.E. and P\'erez, D. (2011).
\newblock {Time and irreversibility in an accelerating universe}.
\newblock {\em International Journal of Modern Physics D}, 20, 1-8.









































































\end{thebibliography}


\end{document}